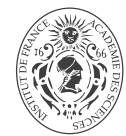

# Electronic structure of 2D van der Waals crystals and heterostructures investigated by spatially- and angle-resolved photoemission

**Irène Cucchi**[a], **Simone Lisi**[a], **Florian Margot**[a], **Hugo Henck**[a], **Anna Tamai**[a] and **Felix Baumberger**[*, a, b]

[a] Department of Quantum Matter Physics, University of Geneva, 24 quai Ernest Ansermet, CH-1211 Geneva, Switzerland

[b] Swiss Light Source, Paul Scherrer Institute, CH-5232 Villigen, Switzerland.

*E-mails:* Irene.Cucchi@ik.me (I. Cucchi), Simone.Lisi@unige.ch (S. Lisi), Florian.Margot@unige.ch (F. Margot), Hugo.Henck@unige.ch (H. Henck), Anna.Tamai@unige.ch (A. Tamai), Felix.Baumberger@unige.ch (F. Baumberger).

**Abstract.**

Angle-resolved photoemission is a direct probe of the momentum-resolved electronic structure and proved influential in the study of bulk crystals with novel electronic properties. Thanks to recent technical advances, this technique can now be applied for the first time for the study of van der Waals heterostructures built by stacking two-dimensional crystals. In this article we will present the current state of the art in angle-resolved photoemission measurements on two-dimensional materials and review this still young field. We will focus in particular on devices similar to those used in transport and optics experiments, including the latest developments on magic-angle twisted bilayer graphene and on the *in-operando* characterization of gate tunable devices.

**Keywords.** nano-ARPES, 2D materials, van der Waals heterostructures, electronic structure.

**Mathematical subject classification (2010).** 00X99.

**Version française abrégée**

La photoémission résolue en angle mesure directement la structure électronique des matériaux et s'est distinguée dans l'étude de cristaux volumiques avec de nouvelles propriétés électroniques. Grâce à des avancées techniques récentes, cette technique peut maintenant être utilisée pour étudier les hétérostructures de van der Waals construites en empilant des cristaux bidimensionnels. Dans cet article, nous présenterons l'état de l'art actuel pour la photoémission résolue en angle appliquée aux matériaux bidimensionnels, et nous dresserons un panorama de ce domaine encore jeune. Nous nous focaliserons en particulier sur des systèmes similaires à ceux utilisés pour des expériences de transport et d'optique, y compris la bicouche de graphène avec une orientation relative entre les couches proche de l'angle magique (*magic-angle twisted bilayer graphene*), et l'effet de la variation *in-situ* de tension dans les hétérostructures de van der Waals.

---

* Corresponding author.

# 1. Introduction

## 1.1. *Context*

The fabrication of single-layer graphene-based field effect devices in 2004 [1] started a new field of research on two-dimensional (2D) materials [2]. The same way graphene was isolated from graphite, 2D crystals have recently been obtained from a multitude of van der Waals (vdW) materials and other layered crystals. Together with advances in the transfer of 2D crystals, this enabled the creation of artificial new materials and novel devices at a pace never seen before.

In bulk form, layered materials suitable for the isolation of 2D crystals have diverse electronic properties and include wide-gap insulators like hexagonal boron nitride (h-BN), semimetals like graphene, or semiconductors like $MoS_2$. In addition, a variety of many-body phases has been reported. Notable examples include high-temperature superconductivity in monolayer $Bi_2Sr_2CaCu_2O_{8+\delta}$ (Bi-2212) [3], nematictiy in FeSe [4, 5], charge and spin-density waves, diverse magnetically ordered states [6, 7] including antiferromagnets with lifted Kramers degeneracy, a Mott state and a putative quantum spin liquid in $TaS_2$ [8, 9] and possibly excitonic insulator phases in $TiSe_2$ [10] and $Ta_2NiSe_5$ [11]. Finally, vdW crystals proved a fertile class of materials for the discovery of topological states. Following the first reports of a strong topological insulator phase in $Bi_2Se_3$, more subtle topologically non-trivial states were reported including a type-II Weyl semimetal phase in $MoTe_2$ [12], a magnetic topological insulator in $MnBi_2Te_4$ [13], topological superconductivity in FeSe [14] and $TaSe_3$ [15] and a dual topology in $Pt_2HgSe_3$ [16].

When thinned down to a single- or few-layers, van der Waals materials often reveal entirely new properties compared to the bulk counterparts. Most famously, transition metal dichalcogenides transform from indirect to direct band gap semiconductors in the monolayer limit [17, 18]. Even more striking is the evolution of $WTe_2$ from a potential type-II Weyl semimetal [19] into a topologically trivial ferroelectric and finally into a robust quantum spin Hall insulator as the thickness is reduced to a single ML [20, 21, 22]. Combining 2D materials with such diverse properties in heterostructures offers exciting prospects for fundamental studies of artificial quantum matter and for applications of such structures in optoelectronic and logic devices or advanced sensors. The mechanical stacking of 2D materials further allows one to control the twist angle between layers. This provides an additional degree of freedom with profound consequences on electronic properties, as exemplified by the recent discovery of superconductivity and correlated insulating states in magic-angle twisted bilayer graphene [23].

To date, the investigation of emergent properties as the thickness of vdW materials is reduced largely relies on transport and optics experiments. Angle-resolved photoemission (ARPES), the most direct probe of the electronic structure, has thus far had a limited impact on the rapidly evolving field of 2D materials. This is primarily because of the significant technical difficulty of such experiments, which the community only begins to master. Yet, progress in the relevant techniques is fast and the potential of ARPES studies in the field of 2D materials is clear. In order to profit from the cross-fertilization between theory and experiment that has long driven progress in quantum matter physics, it is crucial to develop spectroscopic techniques that provide microscopic insight and stringent tests of theoretical approaches. ARPES is very powerful in this regard. It directly measures quasiparticle band structures that serve to benchmark theoretical methods and underlie the interpretation of experiments carried out with other techniques. Moreover, ARPES is intrinsically a many-body spectroscopy and can thus reveal the interactions behind numerous collective phases and fundamental quantities such as the effective mass of interacting particles.

## 1.2. *Angle-resolved photoemission for the study of 2D materials*

ARPES experiments record the kinetic energy and emission angle of photoelectrons emitted from a crystalline sample that is excited by light in the deep ultra-violet (DUV) to soft X-ray range of the spectrum. Photoemission from a crystalline solid does not only conserve energy but also the two momentum components in the surface plane for which translational invariance is preserved. ARPES experiments thus give direct access to the momentum-resolved photocurrent in the solid, which - in the most simple descriptions of the technique - reflects the spectral function [24, 25]. Strong peaks in the latter are usually associated with quasiparticle excitations. The evolution of peak energies with momentum can thus be interpreted as the quasiparticle band structure of the interacting system. In the past, ARPES has been decisive in the characterization of correlated states of matter, most notably in cuprates and iron-based superconductors. In more recent years, photoemission proved equally influential as a probe of the bulk and surface electronic structure of topological insulators, Dirac- and Weyl semimetals [26, 27, 28].

As a momentum space technique, ARPES intrinsically averages over real space. Conventional ARPES experiments are thus performed on single crystals or thin films with hundreds of microns lateral dimensions, which is far beyond the typical size of exfoliated 2D materials [25]. Efficient synchrotron-based ARPES instruments that can map the electronic structure of large samples with high energy and momentum resolution have become increasingly standard in recent years. This enabled studies on 2D materials grown by various *in-situ* deposition techniques. Examples include mono- and few-layer TMDCs on metal surfaces [29] or on graphene/SiC [30, 31, 21, 32, 33], monolayer graphene grown on Ni(111) [34], Ir(111) [35], Cu(111) [36], Ru(001) [37] and on SiC [38, 39, 40, 41, 42], few-layer graphene [43, 44, 45] and graphene/hBN heterobilayers on Cu(111) [46].

Accessing the electronic structure of exfoliated 2D materials by ARPES requires a spatial resolution of the order of 1 $\mu$m. While this is technically challenging, it is still far above the fundamental limit for the spatial resolution of an ARPES experiment given by the uncertainty principle which limits the ultimate momentum space resolution to the inverse of the real space resolution. To date, two main approaches have been pursued to combine real- and momentum-space resolution in an ARPES experiment. The first approach is to use multiple small apertures in a complex electron-optical system to select a small real space area and image a certain energy-momentum window of photoelectrons emitted from this area. While such momentum microscopes or k-PEEMs (photoelectron emission microscopes) often have a large parallel detection range in energy and/or momentum - which improves the efficiency of data collection - they generally require a significant over-illumination of the measured area, resulting in a low overall system efficiency. In addition, the energy resolution of such instruments cannot yet compete with standard hemispherical electron analyzers used for conventional ARPES experiments. The second, more common approach is to use conventional electron spectrometers with large spatial acceptance areas and to achieve spatial resolution by focusing the incoming UV radiation to a small spot [47, 48, 49, 50]. In section 1.3 below, we will discuss this class of instruments in more detail.

## 1.3. *State of the art micro- and nano-ARPES instruments*

Small spot ARPES experiments pose stringent requirements on the light sources. Modern photoelectron spectrometers achieve a momentum resolution better than 1% of a typical Brillouin zone and sub-meV energy resolution. Yet, even in conventional, spatially integrating ARPES experiments, this resolution has to be relaxed to achieve acceptable count rates with the available photon flux. This is particularly true for synchrotron based experiments where the energy resolution of the spectrometer needs to be matched by that of the beamline monochromator, which inevitably reduces the flux on the sample and prolongs acquisition times accordingly. A high photon flux in a narrow bandwidth is thus key for efficient high-resolution ARPES experiments. Spatially resolved ARPES experiments need in addition a small source size and low divergence of the photon beam. Last but not least, ARPES requires DUV or VUV photon energies to access electronic states over the full Brillouin zone of typical materials. Even experiments covering a smaller momentum range require photon energies clearly above the work function of typically 4.5 eV. The combination of these criteria limits the choice of photon sources to high-performance frequency converted lasers and modern undulator beamlines. With the exception of a few truly continuous wave UV lasers, these sources are pulsed, which increases space charge broadening of the spectra. The detrimental effect of the latter is already significant at synchrotrons operated at 500 MHz repetition rate [51] and will likely prove prohibitive for micro-ARPES instruments based on pulsed high-harmonic laser sources, although little data is available to date.

The maximum spatial resolution $\Delta s$ (full width half maximum (FWHM)) that can be achieved with an ideal Gaussian beam is given by the diffraction limit $\Delta s \approx 0.4 \frac{\lambda}{NA}$, where $\lambda$ is the wavelength and $NA$ the numerical aperture. This naturally favors high photon energies. On the other hand, ARPES becomes increasingly challenging at photon energies above ~ 150 eV, principally because the photoemission cross section decreases rapidly with energy. Presently, micro- and nano-ARPES experiments are thus most commonly performed in the range of 6 - 150 eV.

Figure 1 summarizes key-characteristics of typical micro- and nano-ARPES instruments and shows the most common optical schemes currently adopted for focusing the UV beam. At photon energies up to ~ 7 eV, high-quality fused silica diffractive optics can be used. This allows for a simple optical and mechanical design and should permit numerical apertures up to ~ 0.2 and spot sizes down to 400 nm. Existing instruments use smaller numerical apertures though and the highest spatial resolution demonstrated thus far in this photon energy range is ~ 1.8 $\mu$m [54]. Diffractive optics can be used up to the transmission cutoff of LiF near 11 eV. However, currently only spherical lenses are available in LiF which limits the numerical aperture to ~ 0.05 before spherical aberrations become prohibitive.

Laser-based micro-ARPES instruments generally provide a high photon flux in a narrow bandwidth [55, 52]. This enables rapid data acquisition at a high energy resolution. The low photon energy also increases the probing depth, which is often an advantage, and improves the momentum resolution of ARPES experiments. On the other hand, it restricts the field of view in **k**-space, and can lead to dramatic matrix element effects. In addition, the low photoelectron kinetic energy renders laser-based instruments very sensitive to any spurious electric fields near the sample space.

In synchrotron-based ARPES experiments using higher photon energy, the beam can be focused with reflective systems and diffractive Fresnel zone plate (FZP) lenses. The latter currently provide the best spatial resolution. At photon energies around 100 eV, the Maestro nano-ARPES beamline at ALS achieves ~ 120 nm spot size, while 200 nm have been demonstrated at beamline I05 of Diamond Light Source [56]. In the hard X-ray regime, which is presently unsuitable for nano-ARPES, spot sizes below 10 nm were reported [57]. The principal disadvantage of FZPs is the low transmission, typically in the range of $10^{-2} - 10^{-3}$ at the relevant photon energies [47, 57]. This necessitates relaxing the energy resolution in order to achieve acceptable data acquisition rates. Presently, nano-ARPES experiments with FZPs are typically performed at an energy resolution > 50 meV, which is limiting for high-quality samples and prohibits the study of more subtle quantum states. Moreover, increasing the photon bandwidth to boost count rates causes sig-nificant chromatic aberrations, which often dominate the actual spatial resolution of FZP systems [56].

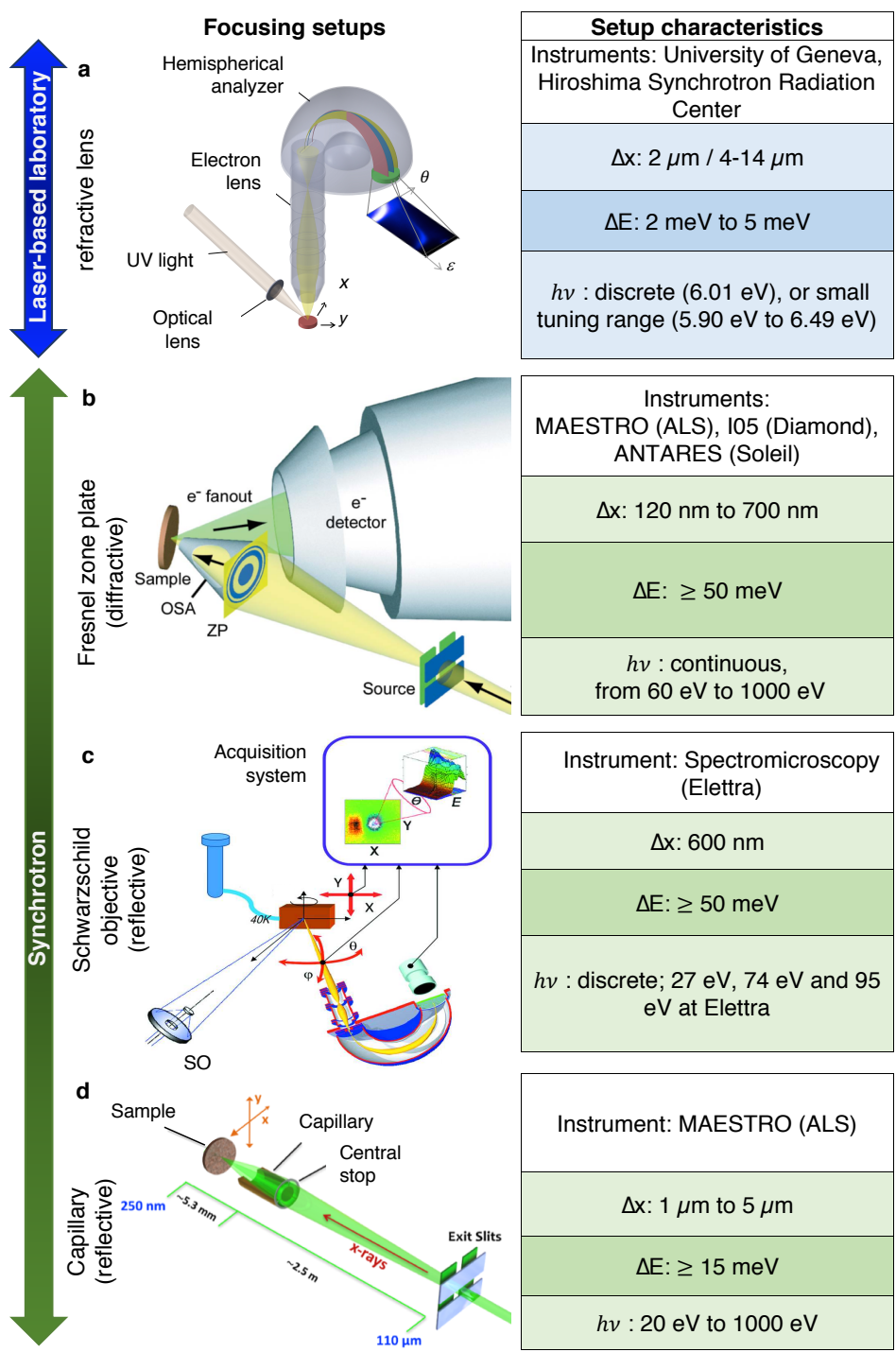

**Figure 1.** Overview of the most common approaches to micro- or nano-ARPES with focused frequency converted laser sources or synchrotron light sources. **a** Laser-based setup with refractive optics (sketch adapted from Ref. [52]). **b** Fresnel zone-plate optics (ZP) with an order sorting aperture (OSA) represented schematically by a cone (Adapted with permission from Kastl *et al.*, ACS Nano 13 (2) [53]. Copyright 2019 American Chemical Society.). **c** Schwarzschild objective (SO) (sketch adapted from Ref. [48]). **d** Elliptical capillary mirror (sketch adapted from Ref. [50]).

Reflective optical systems used for synchrotron based nano-ARPES include Schwarzschild mirrors [48], elliptical capillaries [50] and Wolter-type mirrors [58]. Among these systems, Schwarzschild mirrors have the highest numerical aperture. However, they work at near-normal incidence and thus require multilayer coatings designed for a specific wavelength to achieve good transmission. These coatings reduce the optical quality of the mirrors and limit them to a single wavelength [48]. Elliptical capillaries and Wolter-type mirrors work at small incidence angle and combine good focusing performance with high transmission over a wide range of photon energies [50, 58]. To date, spot sizes of 600 nm (Schwarzschild optics at Elettra [48]), 250 nm (elliptical capillary at Maestro beamline, ALS [50]) and $0.4 \times 4$ $\mu$m (Wolter mirror at Spring-8 [58]) have been reported. These solutions approach the spot size of typical FZP optics [56], with an achromatic design and greatly improved transmission, which are both key for successful ARPES experiments. This has motivated numerous projects to replace or supplement FZPs with reflective focusing optics in small-spot ARPES.

Spatial mapping in nano-ARPES is achieved by mechanically moving the sample. In order to achieve high scanning speed and precision over a large scan range, several nano-ARPES systems use a two-stage design with piezo stick-slip motors for the coarse motion and direct piezo stages for the fine motion [59]. However, for many applications simpler and more robust bellow-sealed sample stages driven by stepper motors mounted on the air-side remain competitive [52, 49].

## 2. First measurements on graphene and on semiconducting transition metal dichalcogenides

### 2.1. *Graphene*

Graphene played a prominent role in the field of ARPES on 2D materials, not only because of its scientific importance but also because it is comparatively easy to measure. Most importantly, high-quality data can be obtained on epitaxial films with conventional, spatially integrating ARPES instruments. Graphene formation from the exposure of hot transition metal surfaces to hydrocarbons [60, 61] was first reported decades before the isolation in 2004 of graphene by mechanical exfoliation. Epitaxial graphene was also among the first materials to be studied by nano-ARPES [62, 63, 44]. The immediate surge in interest in the electronic properties of exfoliated graphene following its mechanical exfoliation motivated efforts to obtain high-quality ARPES data on large-area epitaxially grown graphene. Graphene grown by sublimation of Si from 6H-SiC(0001) in UHV [64, 65] proved particularly suitable to this end [38, 39, 40]. It was later found that the data quality can be further enhanced by hydrogen intercalation, which effectively decouples ML graphene from 4H-SiC(0001) or 6H-SiC(0001) [41, 42]. In line with the limited scope of this brief review, we will not discuss these results here.

### 2.2. *Electronic structure of semiconducting transition metal dichalcogenides*

Early nano-ARPES measurements on exfoliated 2D materials focused largely on transition metal dichalcogenides (TMDCs). These experiments confirmed key-features of the band structure underlying the remarkable properties of these 2D semiconductors. Specifically, they found that the valence band maximum switches from $\Gamma$ in bilayer (2L) and multilayer $MoS_2$ to K in monolayer (1L) $MoS_2$, as shown in figure 2**a** [66]. This change is consistent with the direct band gap in the monolayer, as opposed to indirect band gap in few-layer form, initially deduced from the increased photoluminescence in the monolayer [17, 18]. A further key-aspect in the electronic structure of 2H-TMDCs is the large spin-orbit splitting at the valence band maximum at the K point in monolayer crystals [67]. This splitting has been determined directly with nano-ARPES in exfoliated monolayers of WS$_2$ [68] and WSe$_2$ [69], as displayed in panels **b** and **c**. The broken spin-degeneracy in the K valleys arises from the broken inversion symmetry of the monolayer and is in agreement with the valley-selective excitation with circularly-polarized light observed in optical measurements [70, 71]. Prior to the above nano-ARPES studies, the transition to a direct gap and the spin-orbit splitting at K have already been detected by conventional ARPES on MoSe$_2$ films grown by MBE on graphene/SiC substrates [31].

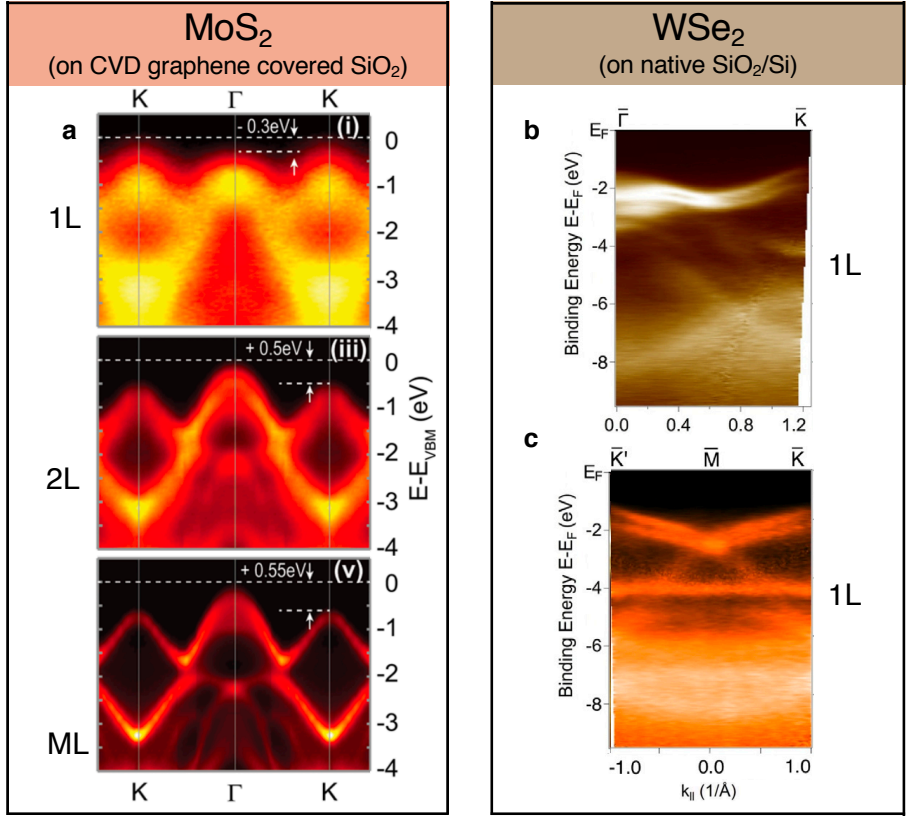

**Figure 2.** ARPES measurements on semiconducting MoS$_2$ [66] and WSe$_2$ [69]. **a** Dispersion along ΓK acquired on 1L, 2L and ML areas of an exfoliated MoS$_2$ crystal. (adapted with permission from Yuan *et al.*, Nano Letters 16 [66]. Copyright 2016 American Chemical Society.). **b,c** Band structure measurements of monolayer WSe$_2$ along ΓK and KMK (adapted with permission from Le *et al.* J. Phys.: Condens. Matter 27 (2015) 182201 [69]. Copyright IOP Publishing. All rights reserved.).

2.3. *Substrate effects*

In freestanding 2D semiconductors, the screening of Coulomb interactions is strongly reduced since the electric field of excess charges extends outside the crystal. This reduced screening has important consequences. It increases the band gap in freestanding 2D semiconductors and causes very large exciton binding energies [72, 73, 74]. In turn screening provides a way to manipulate these properties by embedding the 2D semiconductors in different dielectric environments [75, 76]. Understanding these effects is interesting both for fundamental studies and for potential applications.

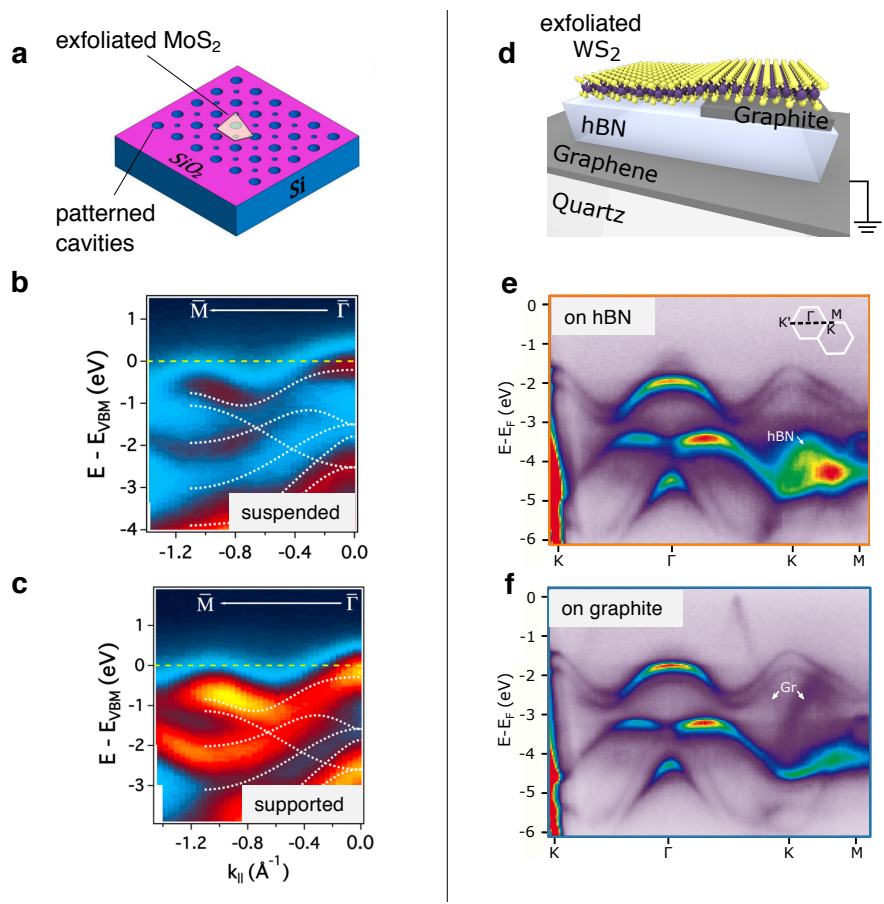

**Figure 3.** Effect of the substrate on monolayer $MoS_2$ [77] and monolayer $WS_2$ [78]. **a** Sketch of the sample prepared by Jin *et al.*. **b**, **c** ARPES measurements for suspended and supported $MoS_2$ (adapted with permission from Jin *et al.*, PRB 91, 121409 (2015). Copyright 2015 by the American Physical Society.). **d** Sample design used by Waldecker *et al.* for photoemission and optical spectroscopy measurements. **e**, **f** $\mu$-ARPES measurements of $WS_2$ on hBN and graphite substrates (adapted figures with permission from Waldecker *et al.*, PRL 123, 206403 (2019). Copyright 2019 by the American Physical Society.).

Nano-ARPES studies investigating substrate induced screening in TMDCs illustrate the potential of this technique to access relevant physics in vdW heterostructures beyond the basic electronic properties of individual 2D crystals. Ref. [77] is an early attempt to identify effects of the substrate on 2D materials with ARPES. In this study, Jin *et al.* exfoliated a monolayer of $MoS_2$ on a $Si/SiO_2$ substrate with 1 $\mu$m deep patterned cavities and measured the electronic structure of freestanding $MoS_2$ and $MoS_2$ in direct contact with the $SiO_2$ substrate, as shown in Figure 3**a**. Establishing a clear conclusion on the substrate interaction from the ARPES spectra proved difficult though, because of the broad spectral features.

A later experiment was performed by Waldecker *et al.* [78]. For this study, the authors used a viscoelastic stamp to transfer an exfoliated monolayer of $WS_2$ and release it at the edge of a graphite flake that partially covered an hBN flake. This allowed a clean comparison of the electronic band structure on substrates with different dielectric properties. Note that both substrates can be recognized on the nano-ARPES measurements shown in figure 3**e,f** (characteristic bands are marked by white arrows). Combining nano-ARPES with complementary optical spectroscopy measurements on the same sample, Waldecker *et al.* showed that the increased screening from graphene leads to a reduction of the quasiparticle band gap of ∼ 180 meV. This reduction occurs via a largely rigid shift of the valence bands, which was attributed to the particular momentum space structure of screening induced by the dielectric environment of the monolayer [78].

## 2.4. *Sample quality*

The short inelastic mean free path of electrons in solids renders ARPES highly surface sensitive [79]. At typical photon energies, ARPES probes the electronic structure in a surface layer of ∼ 0.5 − 2 nm thickness. The technique thus poses stringent requirements on surface cleanliness. On many bulk samples that have been intensely studied by ARPES, already a contamination from adsorbates of less than a percent of a monolayer is sufficient to broaden, if not quench, long lived quasiparticle states [80, 81]. Experiments are thus normally performed in UHV at pressures $< 10^{-10}$ mbar and on sample surfaces that were prepared under comparably clean conditions. Yet, even with these precautions, surfaces often degrade within hours due to residual gas adsorption.

Standard high-resolution ARPES experiments are most commonly performed on bulk crystals cleaved in UHV at low temperature. This is typically done by breaking off a post that was glued on the crystal prior to insertion into the UHV system. This technique is not only simple but also extremely clean and often results in surfaces that are well ordered on the atomic scale. On the other hand, it largely limits such experiments to layered crystals with a natural cleavage plane and is not suitable for the isolation of 2D crystals with controlled thickness or even the fabrication of heterostructures.

The highest quality 2D materials and heterostructures for transport and optical studies are generally obtained by micro-mechanical cleavage combined with the all-dry transfer of exfoliated flakes using viscoelastic stamps. Even if these procedures are performed in a glove box, the exposure of crystals to reactive gases is several orders of magnitude greater than in UHV. Moreover, dry-transfer techniques often leave polymer residues on the surface. The latter can to some extent be removed by solvents and/or "brooming" with the tip of an atomic force microscope. However, such techniques are clearly far from the cleanliness of UHV. Until recently, it was thus not at all evident that ARPES is compatible with sample preparation techniques used in the field of 2D materials.

The contamination of surfaces is a dominant source of spectral broadening in ARPES experiments and frequently obscures relevant physics. Looking at the ARPES data of exfoliated TMDCs in figures 2,3, large differences in the width of spectral features are evident. For instance the data on monolayer $MoS_2$ in figure 2 **a** has line widths of the order of 500 meV, far greater than the spin-splitting at the top of the valence band, whereas the data in figure 3 **e,f** shows line widths around 100 meV. These differences cannot be attributed to the spectrometer resolution. Rather, they highlight sample preparation as a critical aspect of the field.

The consensus today for ARPES studies on 2D materials is to use graphitized SiC, exfoliated hBN or exfoliated graphite as a substrate. This is clearly inspired by the introduction of hBN as a substrate in graphene field-effect devices, which led to an immediate 10-fold increase of the mobility over graphene on $SiO_2$ [82, 83]. Yet, there are still many unknowns and to date, the surface quality of 2D crystals investigated with ARPES is still below that of bulk single crystals cleaved in vacuum. There is thus a clear need for more research in this direction. A key-difficulty is that ARPES rapidly looses its sensitivity if the relevant materials or interfaces are buried. This prohibits the use of hBN encapsulation layers of several nm thickness, which proved suitable for high-quality devices of 2D materials.

Most ARPES studies have been performed on 'open' devices with the surface of the active material exposed. This approach is convenient for graphene and TMDCs for which samples can even be prepared in air and cleaned subsequently by annealing in UHV. Above ∼ 300 °C, the latter also partially desorbs polymer residues, which is often seen to improve the quality of ARPES spectra. However, the use of open devices is challenging for vdW semimetals and metals, which are generally much more reactive, and for air-sensitive 2D semiconductors. Alternative approaches to prepare samples of reactive vdW materials for nano-ARPES will be discussed later in section 4.

## 3. van der Waals heterostructures and Moiré superlattices

Combining 2D materials in heterostructures or twisted homo-bilayers offers unprecedented potential for the design of artificial quantum matter with tailored physical properties. Although only a small subset of possible structures has been explored to date, recent work points to a remarkable variety of properties. TMD-based vdW heterostructures, for instance, host long-lived interlayer excitons [84] and already proved promising for the creation of photovoltaic and optically active devices [85, 86, 87, 88]. Heterostructures of multilayer TMDCs with InSe were used to create a range of robust artificial semiconductors with direct band gaps covering a wide frequency spectrum [89].

The stacking of vdW crystals often creates moiré superlattices. This provides an additional degree of freedom with profound consequences on the physical properties. Most prominently, multiple correlated insulating states and superconducting phases were discovered recently in twisted bilayer graphene [90]. Moiré superlattices arise from the lattice mismatch at the interface and/or from a small twist angle between the layers, but are only stable in systems with strong intra-layer and weak interlayer bonding. They are thus largely unique to vdW heterostructures and are not normally found in heterostructures of oxides or covalently bonded semiconductors.

### 3.1. *TMDC heterostructures*

Band offsets are pivotal for the design of 2D semiconductor-based heterostructures and can be determined directly from ARPES experiments. This is illustrated in figure 4 showing the band dispersion in monolayer $WSe_2$ and in an aligned heterostructure of $MoSe_2$ on $WSe_2$. In the data on $WSe_2$ one can clearly recognize the characteristic spin-splitting at the K-point. On the $MoSe_2$/$WSe_2$ heterostructure, the VBM at the K point of $WSe_2$ is found 300 meV above that of $MoSe_2$ and also above the band extrema of the heterobilayer at Γ where states from the different layers hybridize strongly. The $WSe_2$ K-point thus remains the valence band maximum of the full heterostructure [91]. This implies a type-II band alignment (staggered gap) and is in agreement with earlier first principles calculations [92], although the offset reported in the calculations is slightly larger. The type II band structure is crucial for the existence of interlayer excitons in which photoexcited electrons relax to the conduction band minimum of $MoSe_2$ and photoexcited holes to the valence band maximum of $WSe_2$. Such excitons with photoexcited carriers generated in different layers show remarkably long lifetimes [84]. Offsets between different vdW materials have also been measured with ARPES on $MoS_2$/graphene heterostructures [93, 94].

The highest lying valence band at Γ in monolayer $WSe_2$ (labelled "W") is clearly visible in figure 4 **a**. In the $MoSe_2$/$WSe_2$ heterostructure, two new bands appear at Γ. Besides the "W" edge of $WSe_2$ one recognizes the "M" edge of $MoSe_2$ and another spectral feature, denoted "H".

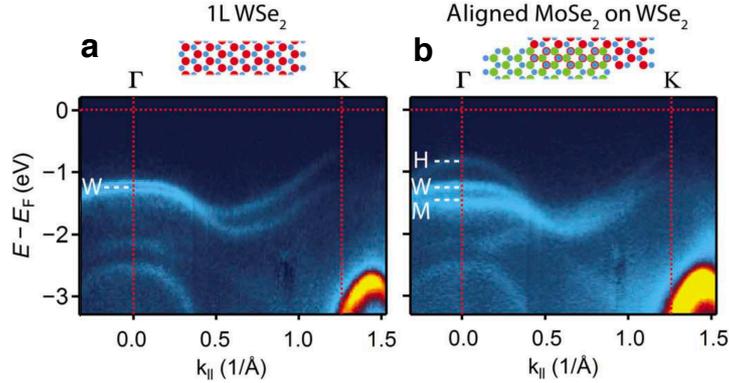

**Figure 4.** ARPES measurement of MoSe$_2$/WSe$_2$ vdW heterostructures. **a** Dispersion in monolayer WSe$_2$. **b** Aligned heterostructure of MoSe$_2$ on WSe$_2$. Reprinted from Wilson *et al. Sci. Adv.* 2017;3: e1601832. Distributed under CC BY.

Based on DFT calculations, this additional spectral feature could be attributed to a signature of commensurate domains present in the heterostructure. MoSe$_2$ and WSe$_2$ share the same crystalline structure and have lattice constants differing by less than 1%. Assuming a rigid lattice, an aligned heterostructure should thus show a long wavelength moiré superlattice. However, such a superlattice inevitably contains areas with unfavorable stacking of the two layers. One thus generally expects commensurate domains separated by domain walls where the strain relaxes. This has previously been reported on aligned graphene/hBN heterostructures with AFM and STM [95].

### 3.2. *Mini-bands in Moiré superlattices*

Moiré super-lattices are common in vdW heterostructures and have been observed at aligned interfaces with a small lattice mismatch between the two materials by STM [96, 97] and conductive AFM [98] . Moiré patterns also appear whenever layers are misaligned by a small twist angle [95, 99, 100]. Despite the modest structural distortion in moiré superlattices, they can have a profound impact at low energy scales. Low-temperature magneto-transport studies on aligned g/hBN heterostructures showed an insulating state with surprisingly large gap, the Hofstadter Butterfly, fractional quantum Hall effect and other novel device characteristics [98, 101, 102].

These phenomena are commonly interpreted as a direct consequence of the formation of mini-bands in the moiré superlattice. Imaging the latter on the relevant energy scales remains a challenge though. Replica bands reflecting the moiré periodicity and mini-gaps opening at certain avoided crossings have been observed by ARPES on numerous systems including aligned graphene/hBN heterostructures [103, 104], graphitized SiC [40, 38] or epitaxially grown graphene on Ir(111) [35] and Ru(0001) [37]. Mini-bands have also been observed on WSe$_2$/graphene heterostructures [91], on bilayer graphene with a twist angle above 5° between the two graphene layers [43], and on MoS$_2$/graphene heterostructures [93, 94]. Recently, anti-crossings due to umklapp processes have also been reported on InSe/graphene and GaSe/graphene heterostructures [105].

These studies illustrate the potential of ARPES as well as its present limitations. The direct observation of replica bands and mini-gaps is clearly an important step towards the controlled band structure engineering in vdW heterostructures. However, present limitations in sample quality and energy resolution of ARPES experiments have so far prohibited a direct comparison of the band structures deduced from these experiments with the unusual magneto-transport properties that triggered much of the interest in moiré systems.

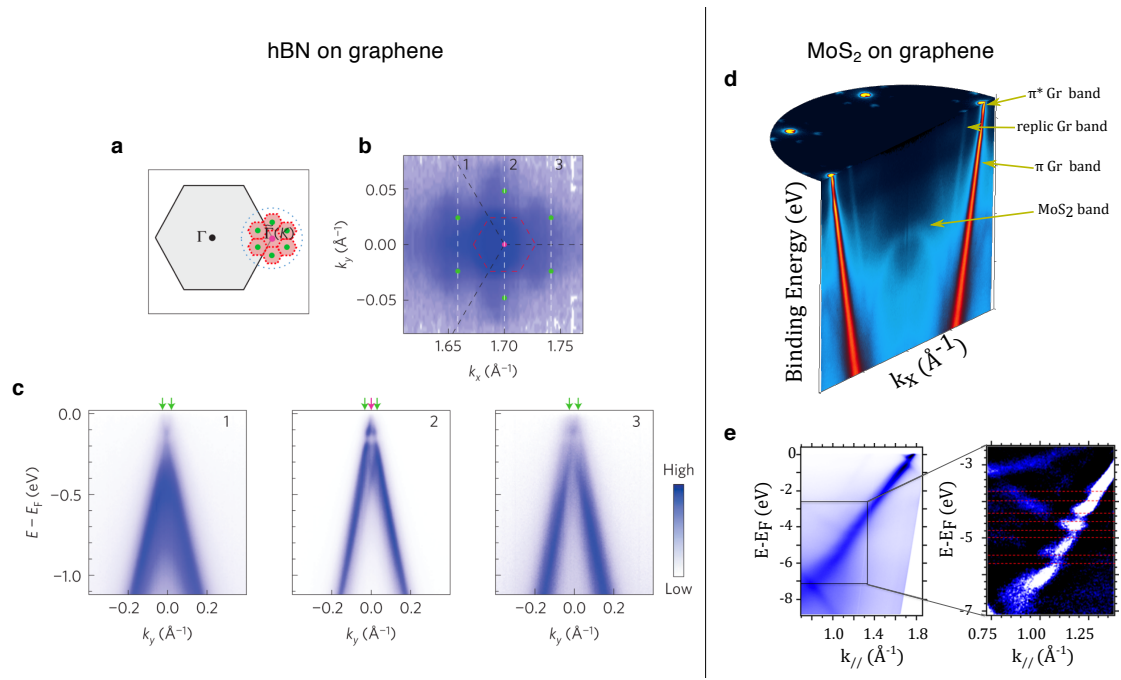

**Figure 5.** Mini-bands in Moiré superstructures. **a-c**: Replica Dirac cones and mini-bands in aligned graphene on hBN. **a** Illustration of the graphene BZ (grey) and the mini-BZ arising from the superlattice potential. **b**: Constant energy map measured at the Fermi level near the graphene K point. **c**: Dispersion along the cuts marked by white dashed lines in panel **b**. Adapted by permission from Wang *et al.* [103], Nature Physics 12 1111 (2016) ), copyright 2016. **d,e**: Superlattice gaps in a $MoS_2$/graphene heterostructure (adapted with permission from Pierucci *et al.*, Nano Letters 16 [94]. Copyright 2016 American Chemical Society.)

### 3.3. *Twisted bilayer graphene*

The mini-bands in moiré superlattices naturally have a small bandwidth, rendering such systems susceptible to correlation effects. In twisted bilayer graphene (TBG) near the so-called *magic angle* of 1.1°, theory predicts the emergence of a half-filled, almost completely flat band that is decoupled from the dispersive states [106]. At integer filling, electrons in such a narrow band might be expected to localize in a Mott-like state. The observation in 2018 of a series of spikes in the resistivity of a TBG device at low carrier density was thus quickly interpreted as a signature of electron-electron correlations [23]. Subsequent experiments also reported superconductivity with a critical temperature up to 3 K and found an even richer phase diagram [90, 107].

Probing the predicted *flat bands* with nano-ARPES is a challenging task and, to date, only two groups reported such experimental data [108, 104]. The fabrication of TBG micro-devices is difficult and involves several low yielding steps. Most importantly, control over the twist angle is limited as graphene flakes frequently rotate by a small angle to relax into a more favorable configuration after being released from the stamp. Moreover, the low throughput of current nano-ARPES beamlines renders measurements on large badges of devices unrealistic. These difficulties are aggravated by the requirement of an open surface. A key step in the fabrication of high-quality TBG devices for transport experiments is the top encapsulation with a thick hBN layer of typically 20-30 nm. This gives the device an increased rigidity and improves the twist angle homogeneity. It also cleanses a large portion of the surface area from contaminants and decouples the TBG from fabrication residues. However, the presence of such a thick layer would completely suppress the photoelectron intensity, and is thus incompatible with nano-ARPES experiments.

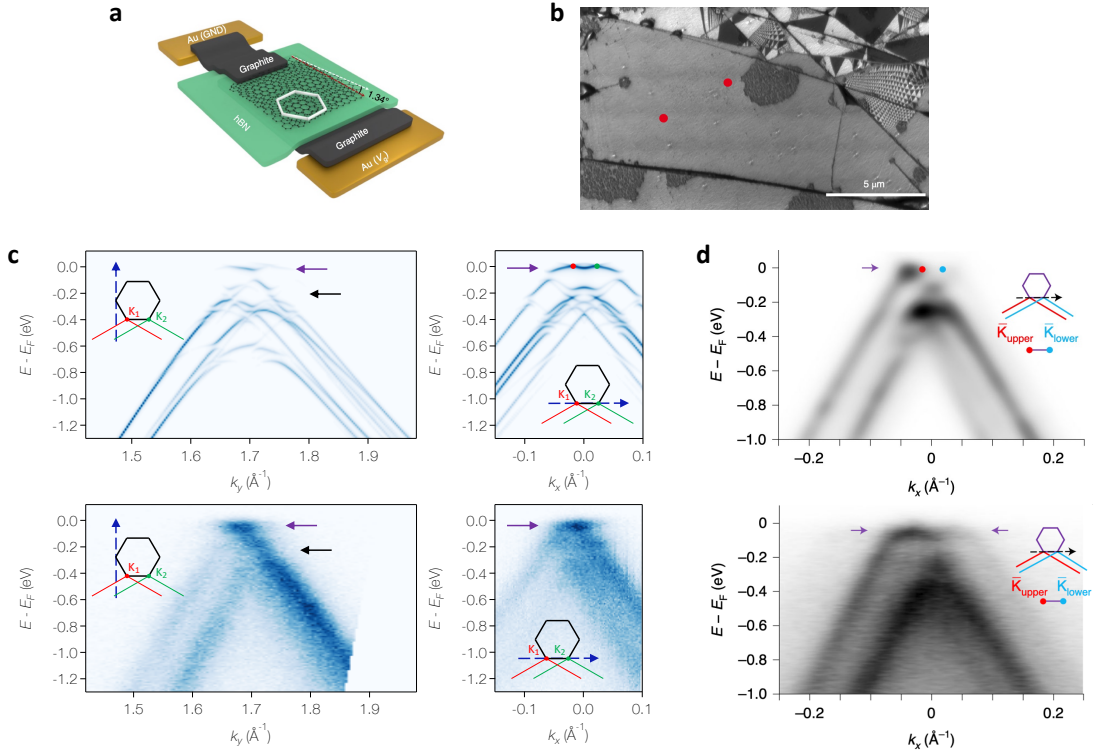

**Figure 6.** Near magic angle twisted bilayer graphene. **a** Sketch of an ARPES-compatible TBG micro-device, with an open architecture as by Lisi *et al.* [108]. **b** Dark field LEEM image of the actual device studied in Ref. [108]. Black and white areas have been identified with AB and BA stacked Bernal bilayer graphene, respectively. Grey areas are TBG with homogeneous twist angle near the magic angle. Triangular reconstruction of low twist angle TBG is visible on the right. Rounded dark grey features are likely polymer agglomerates left after annealing in UHV. **c** Energy-momentum dispersion of TBG reported by Lisi *et al.* for a twist angle of $1.34 \pm 0.03°$ [108]. The calculation (top row) shows spectral weights of a continuum tight-binding model that includes lattice relaxation in the form of a spatial variation of the interlayer coupling parameters [106, 109]. **d** Energy-momentum dispersion reported by Utama *et al.* for a twist angle of $0.96 \pm 0.03°$ [104]. The simulation (top row) is based on an *ab-inito* inspired tight-binding model and includes structural relaxation as well as photoemission matrix elements. The locations in $k$-space of all cuts in panels **c,d** are indicated in the respective insets. The flat bands are highlighted by violet arrows. Red and green dots mark the K-points of the two graphene layers. The black arrow in **c** marks a hybridization gap in the dispersive states. (adapted by permission from Lisi *et al.* [108] Nature Physics, copyright 2020 and Utama *et al.*, [104] Nature Physics, copyright 2020.)

These uncertainties render a thorough characterization of TBG devices - including an inde-pendent twist angle determination - essential to interpret nano-ARPES experiments reliably. In Ref. [108], low energy electron microscopy (LEEM) has been used for imaging large parts of the TBG device prior to nano-ARPES measurements (see figure 6 **b**). Combining bright and dark field LEEM, Lisi *et al.* could identify the most common defects in their open TBG devices and locate large homogeneous areas with twist angles near the magic angle that were later selected for nano-ARPES measurements. An atomic scale structural characterization in the same area was carried out with STM and revealed twist angles varying from 1.31° to 1.37°, sufficiently close to the magic angle value to observe weakly dispersive states. Utama *et al.* used a combination of AFM and microwave impedance imaging after the nano-ARPES experiments to characterize the device used in their study. From these measurements, they determined a twist angle of $0.96 \pm 0.03°$ [104].

Figure 6 **c,d** shows spectral weight calculations and experimental data for selected energy-momentum cuts of near magic-angle TBG reported by Lisi *et al.* [108] and Utama *et al.* [104]. Both experiments reveal the salient features of the TBG band structure predicted by theory. In particular, they show a clear dichotomy of the electronic states with multiple strongly dispersive features at high energy and a nearly flat band at the Fermi level. The former appear from the combined effects of interlayer hopping $t_\perp$ leading to a bilayer splitting of $4t_\perp$ and the superposition in k-space of multiple diffraction replica of the TBG moiré. The interaction of the latter leads to small hybridization gaps, which can just about be detected in some of the data (black arrows in panel **c**). The device studied by Utama *et al.* shows additional diffraction replica arising from the nearly aligned bottom hBN flake, which was found to cause a distinct moiré superlattice with a shorter period than that of TBG [104]. While this complicates the spectral weights of the dispersive bands in their data, it is unlikely to affect the crucial flat band at the chemical potential, which is responsible for the rich transport properties of near magic-angle TBG. The latter is clearly visible in both data sets. However, a reliable determination of its width remains difficult given the resolution of present generation nano-ARPES experiments.

Interestingly, Lisi *et al.* find that the flat band is separated from the dispersive states by a gap of approximately 50 meV, while no such gap is detected in the simulations or the data of Ref. [104] (panel **d**). The authors of the latter study attribute this to the particular twist angle of their device and show that a gap of similar magnitude appears in their calculations if the twist angle is increased to $\approx 1.1°$.

## 4. Reactive materials

The work on TMDCs, graphene and hBN discussed in sections 2 and 3 used 2D crystals exfoliated and stacked in air. This approach greatly simplifies device fabrication but is applicable only because these materials do not oxidize or otherwise degrade irreversibly when exposed to ambient conditions. Most adsorbed species can then be removed effectively by simple vacuum annealing. The vast majority of vdW materials, however, reacts more or less strongly with oxygen and water and often degrades irreversibly within seconds or minutes when exposed to air. The study of such materials by nano-ARPES has long been a challenge. Very recently, however, two groups reported successful measurements with a sample design inspired by advances in device fabrication pioneered in transport experiments on 2D materials.

The introduction of hBN as a substrate in devices significantly improved the carrier mobility in graphene [82]. Shortly thereafter, it was found that adding an hBN capping layer enhances the mobility even further and renders devices much more stable [110, 111]. This improvement was attributed to an active self-cleansing effect, in which the contaminants trapped during the assembly are separated into isolated pockets, leaving large atomically clean interfaces [112, 113]. In addition, hBN encapsulation reduces the effects of unwanted polymer contamination because the active material is picked up using vdW forces between 2D materials rather than by direct contact with a polymer coated stamp. Fabricating encapsulated samples in the protective atmosphere of a glove box was later found to provide high quality transport data from reactive vdW materials even if the devices were exposed to air in between fabrication and measurements [20, 22, 6].

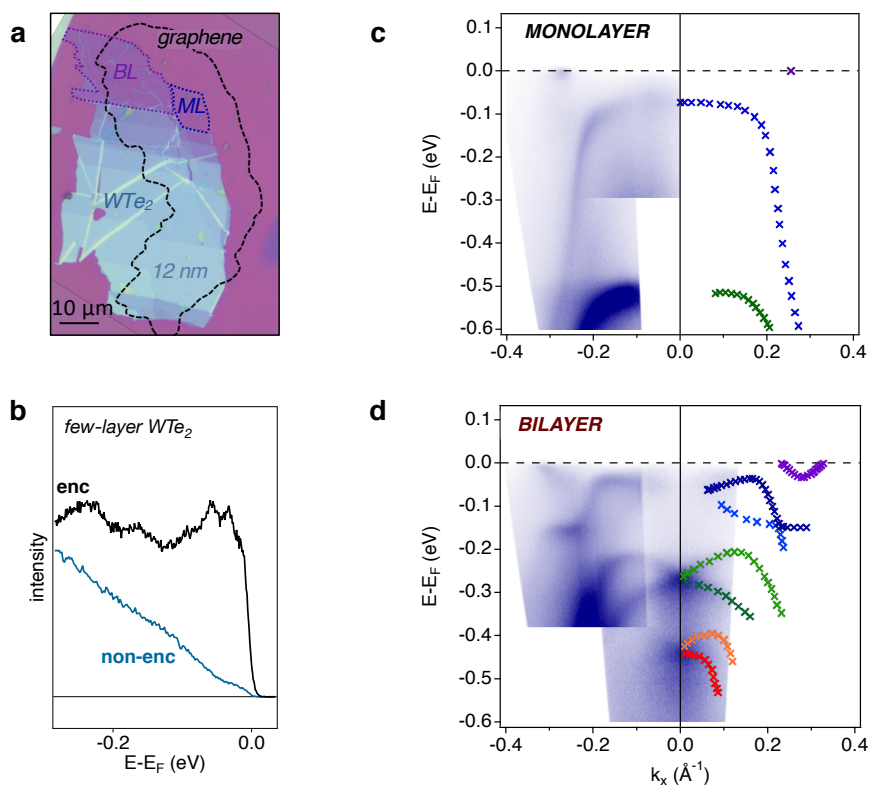

**Figure 7.** Laser $\mu$-ARPES data from exfoliated $T_d$-WTe$_2$. **a** Optical micrograph of the WTe$_2$ 2D crystal encapsulated with monolayer graphene (dashed black contour). **b** ARPES spectra near normal emission from encapsulated and exposed multilayer WTe$_2$ taken on the structure shown in **a**. **c,d** $\mu$-ARPES data from the encapsulated monolayer and bilayer area of the WTe$_2$ flake marked in **a**. Colored symbols are band positions extracted from multiple data sets. (Panels **c,d** adapted with permission from Cucchi *et al.*, Nano Letters 19 [52]. Copyright 2019 American Chemical Society.)

Encapsulation in a glove box was also used in $\mu$-ARPES experiments and proved effective for the protection of air-sensitive 2D crystals [52, 114]. However, unlike in transport experiments, limiting the thickness of the encapsulating layer is essential for ARPES. For a typical inelastic mean free path of $\sim 5$ Å at the photon energies used for synchrotron based nano-ARPES, the encapsulation with a single monolayer of graphene will already reduce the signal by $\sim 50\%$. Encapsulation below 5 nm of hBN, as it is common for transport devices, would lower the count rate of the active material by more than $10^4$, rendering ARPES experiments unfeasible. It is thus imperative to reduce the thickness of the encapsulation layer as much as possible and ideally to a single monolayer. Unfortunately though, monolayer hBN on polymer stamps is nearly invisible in a conventional optical microscope, which makes it very demanding to handle. The authors of Refs. [52, 114] thus both opted for encapsulation by monolayer graphene.

The sample used by Cucchi *et al.* for the investigation of $T_d$-WTe$_2$ is shown in Fig. 7 **a**. Note that the WTe$_2$ flake (turquoise) was only partially encapsulated by graphene, which allowed the authors to directly demonstrate the protective function of the encapsulation layer, as illustrated in figure 7 **b** [52].

Bulk WTe$_2$ in its orthorhombic, inversion symmetry broken T$_d$ phase, is a semi-metal with record high magnetoresistance [115], which was attributed to an exceptionally precise compensation of electron and hole-like carriers [116]. In 2015, WTe$_2$ was also predicted to be a novel form of Weyl semimetals, dubbed "type-II", that breaks the Lorentz invariance of electrodynamics [19]. While other type-II Weyl semimetal were confirmed in the meantime, the prediction could not be verified to date in the case of WTe$_2$ [117, 118]. T$_d$-WTe$_2$ attracted the interest of the 2D materials community when Qian *et al.* suggested a $p-d$ band inversion in monolayer crystals under strain [119]. Such a band inversion would drive WTe$_2$ into a quantum spin Hall insulating phase with topologically protected edge states. Early experiments on monolayer WTe$_2$, however, suffered from its high reactivity [120] and reliable transport data down to the monolayer limit became available only several years after the prediction of the quantum spin Hall state. These experiments did not only show that WTe$_2$ is a small gap semiconductor even in the absence of strain, but also reported robust edge conduction in the monolayer [20, 22].

The ARPES data of Cucchi *et al.* could clearly resolve a gap of $66 \pm 7$ meV in monolayer WTe$_2$ [52]. A comparison with DFT calculations (not shown in Fig. 7) further allowed for a reliable identification of the orbital character of the relevant states, which confirmed the topological band inversion underlying the quantum spin Hall phase. The same authors also studied bilayer WTe$_2$ (Fig. 7 **d**) and found that its electronic structure is remarkably different from the monolayer. Most importantly, it shows the hallmarks of a strong Rashba splitting in the valence bands with a lifting of the spin degeneracy away from the time reversal invariant Γ point [52].

Before the work on WTe$_2$ and InSe from Refs. [52, 114], direct electronic structure measurements on air-sensitive 2D materials were restricted to thin films grown *in-situ* by various deposition techniques [121, 122, 32, 33, 21]. While this continues to be an important method, it is significantly less flexible than exfoliation. Moreover, it is often challenging to obtain high-quality films. Epitaxially grown 2D crystals often have multiple domains with different azimuthal orientations and/or different thickness and lateral dimensions as small as a few nm [123]. This is also the case in WTe$_2$ as shown by the topographic STM image in figure 8**b** [21], where domain sizes are well below the resolution of present-generation nano-ARPES instruments, which inevitably results in a certain averaging of spectra from different domains. Moreover, the chemical purity and stoichiometry of thin films is frequently below what can be achieved in exfoliated crystals. On the other hand, even in state-of-the-art glove boxes, samples are exposed to far higher partial pressures of reactive gases during the fabrication process when compared with thin-film deposition in UHV. In addition, the chemicals needed to remove polymer residues after the dry-transfer might leave residues on the surface themselves. It is thus not a priori clear which approach will produce higher quality electronic structure data.

To date, WTe$_2$ is the only reactive vdW material for which ARPES data from an *in-situ* deposited thin film and an exfoliated crystal exists. The direct comparison of these data in figure 8**c** shows that the quality, as measured by the line width near the Fermi level, is clearly superior for exfoliated and encapsulated crystals [52, 21]. Further studies will need to show whether this is an isolated case or a general trend. We also remark that the conventional synchrotron ARPES spectrum from MBE grown WTe$_2$ is additionally broadened by the instrumental resolution, whereas the latter is negligible in the laser $\mu$-ARPES data of Ref. [52].

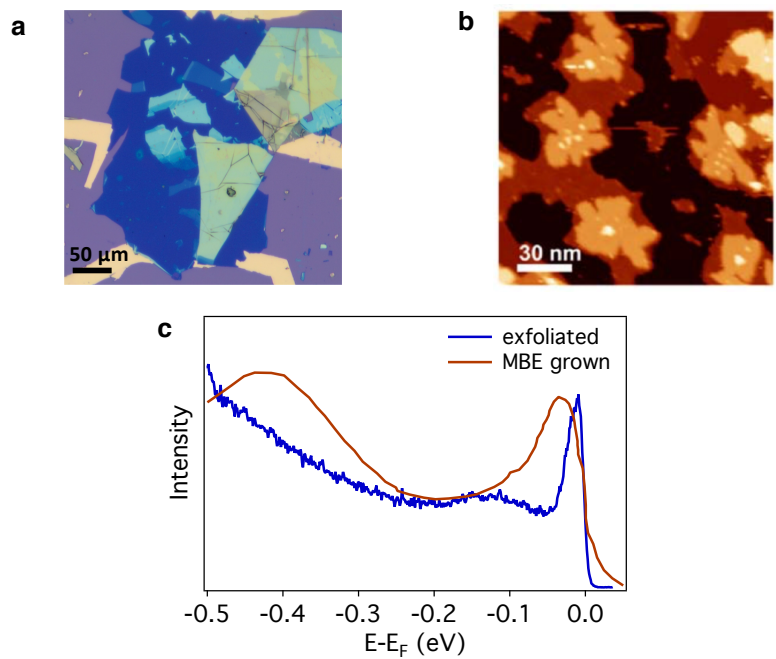

**Figure 8.** Comparison of ARPES data from exfoliated and MBE grown monolayer WTe$_2$. **a** Optical micrograph of the encapsulated WTe$_2$ crystal measured by Cucchi *et al.* [52]. **b** STM topographic image of MBE grown WTe$_2$ on SiC [21]. Different colors correspond to monolayer, bilayer and trilayer WTe$_2$. Adapted by permission from Nature Physics (Quantum spin Hall state in monolayer 1T'-WTe$_2$, Tang *et al.* [21]), copyright 2017. **c** Direct comparison of spectra at the valence band minimum from the exfoliated sample and the MBE grown thin film in **a,b**.

## 5. Tuning the carrier density in 2D crystals

### 5.1. *Near surface doping by alkali deposition*

Field effect tuning of physical properties is fundamental for the study of 2D materials. The most common approach to simulate electric gate fields in ARPES experiments is *in-situ* deposition of alkali atoms. In an idealized picture, alkali deposition produces an ionic adlayer on the sample surface that is screened by electron accumulation in the 2D material. Alkali deposition should thus be closely related to electrolyte gating [124]. The evolution of the electronic structure following alkali deposition has been extensively studied on graphene [125, 40, 38], monolayer MoS$_2$ [126] and monolayer WS$_2$ [127]. These experiments confirmed the direct band gap in monolayer WS$_2$ and found a gap renormalization and increased spin-orbit splitting in the K valleys at high doping, which was attributed to the formation of trions [127].

Surface doping with alkali presents, however, major drawbacks. First, this method is generally irreversible. Even annealing at high temperature in UHV cannot fully remove the adsorbed atoms, and annealing cycles can induce other types of disorder on the sample [128]. As a consequence, *in-situ* fine tuning of the carrier density with alkali is very difficult. In addition, charged impurities on the sample surface are known to affect transport properties [128, 129], which complicates the comparison between a sample doped by adsorbed alkali ions and an actual device in operation. Finally, alkali atoms frequently diffuse to subsurface sites [130], which can alter the structure and intertwine electrostatic effects and chemical doping.

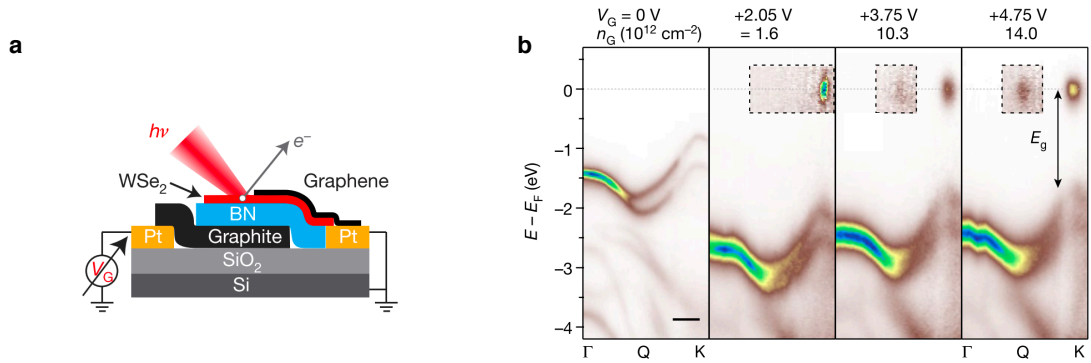

**Figure 9.** *In-operando* ARPES measurements of gated 2D materials. **a**: Sketch of the vdW heterostructure for *in-situ* electrostatic gating used by Nguyen *et al.* [131]. **b**: Band dispersion in WSe$_2$ along ΓK for increasing gate voltage. Gate voltages and carrier densities are indicated on top of each spectrum. Adapted by permission from Nguyen *et al.* Nature 572, copyright 2019 [131].

### 5.2. *Reversible in-situ electrostatic gating*

*In-operando* ARPES studies of gated 2D materials have much potential for providing deep insight into the physics of these fascinating systems. One of the first successful experiments of the kind is shown in figure 9. Nguyen *et al.* used a heterostructure with a graphite gate electrode isolated by a hBN gate dielectric from different TMDC semiconductors. The particular device shown in figure 9**a** used a WSe$_2$ flake containing a monolayer, bilayer and trilayer area. This flake was contacted by a graphene monolayer connected to a Pt electrode that covered most of the wafer surface. This design permits substantial electric fields with low gate voltages and minimizes stray electric fields that might perturb the electron optics of the ARPES spectrometer.

The direct gate-field dependent measurements of the quasiparticle band structure reported in this study confirmed key electronic properties of TMDCs. In particular, they demonstrated that the conduction band minimum in monolayer WSe$_2$, WS$_2$, MoSe$_2$ and MoS$_2$ is indeed at the K-point, resulting in a direct band gap. At high gate fields, Nguyen *et al.* could also populate the Q valley in WSe$_2$ and demonstrate directly that its energy is slightly above the conduction band minimum at K. They further showed that this situation reverses in bi- and trilayer WSe$_2$ where the Q valley dips below the K valley, resulting in an indirect gap, consistent with prior results from optics [17]. Near simultaneous with the work of Nguyen *et al.*, Joucken *et al.* demonstrated *in situ* electrostatic gating of Bernal bilayer graphene during ARPES measurements [132]. Subsequently, *in-operando* ARPES studies have also been reported on monolayer graphene [133, 134, 135] and bilayer graphene with large twist angle [136, 137].

The study of Nguyen *et al.* also discussed specific difficulties of *in-operando* ARPES experiments. Most importantly, ARPES is performed in a dynamic equilibrium in which the photocurrent is continuously compensated through different channels. From their systematic gate voltage dependence, Nguyen *et al.* conclude that at low gate voltages, the photocurrent is compensated predominantly via vertical charge transport through the photoexcited hBN. A significant photo conductivity of hBN in ARPES experiments is also evident from the measurements of thick hBN flakes in Ref. [138]. At high positive or negative gate voltage, Nguyen *et al.* find a significant lateral band bending and current flow to the graphene electrode in contact with the TMDC flake. This band bending causes a significant variation of the electrostatic potential over the dimension of the beam in the nano-ARPES instrument, resulting in a strong broadening of spectral features.

Devising methods to reduce this lateral band bending will be essential to unlock the full potential of *in-operando* ARPES measurements of gapped systems.

## 6. Conclusion

Recent technical developments improved the spatial resolution of ARPES experiments to the submicron regime. Combining this instrumental advance with device fabrication techniques developed in the field of 2D materials opens exciting perspectives for unravelling fascinating physics. Nano-ARPES on 2D materials is still a young field. However, its potential is already evident from a number of successful recent studies discussed in this brief review investigating systems ranging from semiconducting TMDCs to semimetals, topological systems and complex heterostructures including twisted bilayer graphene. Particularly exciting for future developments are *in-operando* experiments under applied gate fields. Extending such studies from graphene and semiconducting TMDCs to topological systems and phase transitions in interacting systems promises unique insight into the rich physics of 2D materials.

However, more research is needed to unlock the full potential of nano-ARPES for the study of 2D materials. Two areas stand out in this regard. The energy resolution of nano-ARPES instruments needs to improve from presently ∼ 50 meV to the few meV range, as is already standard in conventional, spatially integrated ARPES. Such an improvement appears feasible thanks to recent advances in reflective optics. Particularly promising is the use of focusing capillaries, which was pioneered at ALS and is set to become the new standard for synchrotron based instruments. Equally important is work devoted to reducing contaminations and other imperfections in devices. Despite the only modest energy resolution of current generation synchrotron nano-ARPES instruments, the data quality in most recent nano-ARPES studies is still limited by the sample. This is particularly true for gated devices.

## 7. Acknowledgement

This work was supported by the Swiss National Science Foundation (SNSF). We gratefully acknowledge discussions with F. Bruno, E. Cappelli, I. Gutiérrez-Lezama, A. Hunter, J. Issing, S. McKeown Walker, A. Morpurgo, and N. Ubrig.